\documentclass[conference]{IEEEtran}
\IEEEoverridecommandlockouts
\usepackage{cite}
\usepackage{amsmath,amssymb,amsfonts}
\usepackage{minted}
\usepackage{algorithmic}
\usepackage{graphicx}
\usepackage{textcomp}
\usepackage{xcolor}
\usepackage[hyphens]{url}
\usepackage{hyperref}
\usepackage{booktabs}
\usepackage{makecell}
\usepackage{multirow}
\def\BibTeX{{\rm B\kern-.05em{\sc i\kern-.025em b}\kern-.08em
    T\kern-.1667em\lower.7ex\hbox{E}\kern-.125emX}}

\begin{document}

\title{
\footnotesize
\framebox[1.01\width]{\parbox{\dimexpr\linewidth-2\fboxsep-2\fboxrule}{If you cite this paper, please use this reference: A. Lavandier, B. Buil, C. Gaber, E. Baccelli. \emph{treVM: Tiny Rust Embedded Virtual Machines with WASM on Variable Resource-Constrained Hardware}. Proceedings of IEEE DCOSS-IoT, 2026.}}
 \ \\ \ \\ \ \\
\Huge
treVM: Tiny Rust Embedded Virtual Machines with WASM on Variable Resource-Constrained Hardware
}

\author{\IEEEauthorblockN{Antoine Lavandier}
\IEEEauthorblockA{\textit{Inria}}
\and
\IEEEauthorblockN{Bastien Buil}
\IEEEauthorblockA{\textit{Orange Research \& Cnam}}
\and
\IEEEauthorblockN{Chrystel Gaber}
\IEEEauthorblockA{\textit{Orange Research}}
\and
\IEEEauthorblockN{Emmanuel Baccelli~\footnote{Corresponding author: emmanuel.baccelli@inria.fr}}
\IEEEauthorblockA{\textit{Inria \& FU Berlin}}
}

\maketitle

\begin{abstract}
Software stacks embedded on microcontroller-based hardware typically provide rudimentary APIs programmed in C/C++, basic connectivity and, sometimes, a firmware update mechanism. 
Such coarse mechanisms contrast with widely used APIs and more advanced networked interaction expected from software stacks deployed on less resource-constrained hardware (microprocessor-based).
In this paper, we aim to bridge this gap by designing  
treVM, a generic scheme to host high-level WebAssembly code capsules, 
bolted on a general-purpose Rust embedded software platform, able to run on a large variety of 32-bit microcontrollers. Not only can treVM capsules host highly customizable business logic, but capsules can also be securely updated on demand over the network, on devices already deployed in the field. We implement treVM in Rust, on top of Ariel OS, a general-purpose RTOS, and we publish the code as open source. Based on our implementation, we validate the feasibility of treVM on commonly available boards, and we report on extensive benchmarks we performed on heterogeneous hardware including Arm Cortex-M, RISC-V, and Xtensa microcontroller architectures. As such, treVM provides a promising new framework to secure continuous deployment of embedded software on low-power networked devices.
\end{abstract}


\section{Introduction}

A microcontroller unit (MCU) is the central component for billions of
machines used on a daily basis, such as sensors and actuators of various types. These are connected at the 
edge of networks supporting distributed systems in heterogeneous segments --
spanning from smart home to automotive, and from
factory floors to precision agriculture or even fleets of small satellites.

While continuous deployment (CD) has become a \emph{must} for most software running on networked machines,
CD remains a bottleneck for software embedded on MCUs. 
Indeed, to this day, software updates for MCUs often lack convenient and secure partial update mechanisms over the network (a.k.a. over-the-air, or OTA~\cite{el2022mcu-ota-secure-survey}).
Researchers have studied and proposed different mechanisms for partial updates, such as in~\cite{wei2024-intermittent-OTA,kwon2021-partial-firmware}. In the field, however,
if MCU software updates happen at all, these remain predominantly monolithic (so called \emph{firmware} updates). 

On the one hand, however, software on microcontrollers is often complex enough to yield the use of an operating system which takes care of common embedded systems chores, upon which high-level "business" logic can be based. Various open source OS in this space are available, surveyed for instance in \cite{hahm2015operating}, also known as real-time operating systems (RTOS). Traditionally in C/C++, a number of newer RTOS are written in Rust such as Tock OS~\cite{levy2017tock-os}, Embassy~\cite{embassy} or Ariel OS~\cite{frank2025ariel}. The main appeal of Rust compared to C/C++ for systems software is the memory safety it delivers~\cite{google-rust-blog}, and modernized tooling expected to facilitate matching next-level safety and security requirements. 
On the other hand, customizable high-level logic can be partitioned and updated independently of the lower-level OS logic in various ways. A common technique is to encapsulate business logic in small virtual machines. For example, prior research work such as~\cite{zandberg2022femto,baccelli2018scripting} or industrial products such as MicroEJ~\cite{microEJ} leverage small VMs based on eBPF, JavaScript or Java, interfacing with an RTOS via system calls. A growing trend in this domain is to use WebAssembly (Wasm) virtual machines, such as in ~\cite{skrivankova2025smart,buil2025tinyml,van2025cyberphysical-wasm}. The main appeal of Wasm in this field is the cross-hardware application code portability it provides, and the isolation it promises. 

Nevertheless, the use of both Rust and Wasm is rather new in this context. For instance, in~\cite{zandberg2022femto}, authors highlight the overhead of using Wasm virtualization in terms of memory footprint compared to alternative virtualization techniques or native execution. On the other hand, in~\cite{ayers2022tighten-rust}, authors analyze the relatively bigger size of binaries with Rust. Thus, more generally, in the following, we aim to provide a reality check. Namely: what performance should we expect when using small Wasm runtimes, hosted on a Rust embedded software platform, running on current microcontrollers?

\subsection{Paper Contributions}
\begin{itemize}
    \item We survey and benchmark the performance of various Wasm runtimes on different types of microcontrollers including Cortex-M, ESP32 and RISC-V;
    \item We design \emph{treVM}, a generic scheme to host Wasm code snippets on top of heterogeneous MCUs, bolted on top of a common Rust embedded RTOS, whereby Wasm code snippets can be updated securely over the network, independently of the RTOS;
    \item We implement a PoC using \emph{treVM} to securely update logic of different types of WebAssembly code snippets, which we thereafter refer to as \emph{capsules}, which can run on heterogeneous microcontrollers (Cortex-M, ESP32 and RISC-V);
    \item We measure performance of treVM on a popular board based on a Cortex-M microcontroller in terms of execution speed, memory size, network transfer costs, and discuss these results.
\end{itemize}
\section{Related Work}

Recent work such as \cite{zhang2025research}, \cite{Moron2025benchmark} survey Wasm runtimes.
A number of other studies analyze the performance of resource-constrained Wasm runtimes. For instance, in \cite{Moron2025benchmark}, authors compare the performance of WebAssembly,  MicroPython, and Lua on an STM32L4R9-DISCO board. However, prominent runtimes such as Wasmtime are not included in this comparison. 
In \cite{zandberg2022femto}, a  performance evaluation compares the WASM3 runtime against eBPF and microPython for fletcher32 logic on an Arm Cortex-M microcontroller but does not consider any other Wasm runtime.  The article \cite{Wang2022} compares the performance of standalone WebAssembly runtimes (\emph{i.e.} not web-based) against native execution without considering embedded devices.
Meanwhile, \cite{Hilbig2021} study WebAssembly binaries, but unfortunately not the runtimes themselves, and do not include any detailed performance analysis.
Spies \emph{et al.}~\cite{Spies2021} comparatively evaluate standalone WebAssembly against web-based WebAssembly and native code, but focus their evaluation on x86-64 microprocessors with Linux.
Gackstatter \emph{et al.} \cite{Gackstatter2022} study the performance of WebAssembly for small serverless Function-as-a-Service (FaaS), but focus on the microprocessor segment, instead of microcontrollers.
Zhang \emph{et al.}~\cite{zhang2025research} compare Wasm runtimes but do not measure performance, and do not focus on microcontrollers.

Other prior works focus on extending Wasm runtimes to add various functionalities such as remote management~\cite{ribeiroWASMICOMicrocontainersMicrocontrollers2024}, peripheral APIs~\cite{liWiProgWebAssemblybasedApproach2021}, partial code offloading~\cite{liWiProgWebAssemblybasedApproach2021}, fine-grained modules permissions~\cite{liuAerogelLightweightAccess2021}, and controls on energy consumption \cite{liuAerogelLightweightAccess2021} and memory usage~\cite{liuAerogelLightweightAccess2021,ribeiroWASMICOMicrocontainersMicrocontrollers2024}.

In a nutshell, there is so far no encompassing study considering both a wide variety of microcontrollers, and newer Wasm runtimes applicable in this space. Furthermore, no available study has focused on Rust embedded platforms. 
\section{Wasm Subsystem}

\subsection{Surveying Available Wasm Runtimes} \label{runtimes survey}

Properly choosing a runtime is vital to benefit from the advantages of WebAssembly, as it influences the speed, safety and memory footprint of executing the software capsules. 
Recent works such as \cite{zhang2025research}, \cite{Moron2025benchmark} survey several Wasm runtimes. Building on this foundation, our benchmarking aims to evaluate these runtimes in order to facilitate choices tailored to specific embedded applications. The following runtimes stand out as noteworthy options. \

Bytecode Alliance, a nonprofit involved with the standardization of WebAssembly and WASI (WebAssembly System Interface), distributes two WebAssembly runtimes: \textbf{Wasmtime}, their flagship Wasm runtime, which supports, since mid 2024, embedded targets through its Pulley interpreter which uses custom intermediate bytecode, and WebAssembly Micro Runtime (\textbf{WAMR})~\cite{wamr}, their lightweight runtime in C, which supports embedded target by plugging onto bare-metal OS like Zephyr, Free RTOS or RIOT \cite{riot-wrappers,baccelli2018riot}. 

Other runtimes have this quality such as \textbf{Wasmi}~\cite{wasmi} which is used on constrained devices and for blockchain applications. It is written in Rust and actively maintained. \textbf{Wasm3}~\cite{wasm3} is another  runtime explicitly targeting embedded devices. It does not have any dependencies but is however unmaintained since 2022.

\textbf{Wizard}~\cite{FastInPlace2022} is an in-place interpreter written in Virgil which supports multiple targets like x86, jar, and Wasm, but does not support microcontrollers.

\textbf{Wasm-interpreter}~\cite{wasm-interpreter} is a minimal in-place Wasm interpreter written in Rust by the German Aerospace Center and aimed at safety applications. It implements a fast in-place interpreter described in the Wizard paper~\cite{FastInPlace2022}. 

\textbf{Wasefire}~\cite{cretinWasefire2026} is a security framework to execute applets on microcontrollers that provide an in-place interpreter, while the overall framework currently only supports nRF52840 and an FPGA, the interpreter can be run on all of the common embedded architectures.





\subsection{Candidate Wasm Runtimes Selection}
We selected a few candidate runtimes for our Wasm subsystem micro-benchmarks based on the following criteria (i) runtimes must be open-source to facilitate reproducibility and customization of our results, (ii) runtimes must be actively maintained, (iii) runtimes must target embedded explicitly, i.e. common microcontroller hardware architectures such as ARM-32, RISCV-32, and Xtensa, (iv)~runtimes must provide portable executable bytecode. 

We purposely exclude native ahead-of-time (AOT) compilation such as the one proposed by WAMR as this method produces code tailored too specifically to the underlying MCU architecture, and hence limits the portability of Wasm code. Note, however, that this does not exclude the use of bytecode derived from WebAssembly as long as it stays portable between our target architectures without further work. 

Based on the above criteria, we selected WAMR, Wasmi, Wasmtime, Wasm-interpreter and Wasefire as well as some of their variants for performance comparison, as described next.

\subsection{Runtime Evaluation Methodology}
\label{sec-setup}
\subsubsection*{Micro-benchmarks Selection}
We used two suites of benchmarks for our evaluation, both of which are well known suites in this context. The first is CoreMark \cite{coremark2012}, a benchmark aimed at describing the performance and abilities of CPUs that produces a single number that makes qualitative comparisons easy. The second is Embench \cite{patterson2019embench}, a suite that specifically targets embedded computing. It features various code snippets representing typical applications used in constrained contexts, each providing a single numerical score, which allows for more granular evaluation.

Note that, as those benchmarks are written in C, we had to find (or otherwise produce) WebAssembly versions of those benchmarks. For CoreMark, we relied on the Wasm port done by  Wasm3~\cite{Wasm3Wasmcoremark2026}. In the case of the Embench suite, we couldn't find an existing WebAssembly port so we created one by compiling each benchmark separately using the Emscripten C-to-Wasm compiler. The WebAssembly versions of the Embench suite that we produced are part of the open-source contributions of this work. 



\subsubsection*{Hardware, Toolchain and Software Setup}
In order to be as comprehensive as possible, we ran the above benchmarks on four different board/MCU combinations featuring three different architectures and three different vendors: the Raspberry Pico 2 W board featuring a RP2350 (Arm Cortex-M33 ThumbV8M), the nRF52840DK board (Arm Cortex-M4, nRF52840 MCU, Arm ThumbV7), the ESP-WROOM-32 (Espressif Xtensa) and the ESP32-C6 (RISC-V 32-bit). 

Our evaluation benchmarks were implemented on top of a common embedded software platform: Ariel OS \cite{frank2025ariel}. This operating system, written entirely in Rust, supports all the hardware mentioned above.

The binaries were built using the nightly rust compiler \texttt{rustc 1.93.0-nightly (b84478a1c 2025-11-30)} and optimized for binary size. The Pulley bytecode conversion of the Wasm base capsules was also optimized for size using Wasmtime version 38.0.3.

\subsubsection*{Methodology} We use the CoreMark benchmark in two ways. First, we get an overview of comparative performance using its score and second, we measure the static memory usage of each runtime. We define the latter as the sum of the \texttt{.text}, \texttt{.rodata} and \texttt{.data} sections of the ELF binary minus the size of the capsule. This leaves only the contributions to code size coming from the operating system and from the measured runtime. Since both are mostly independent of the specific capsule code, we did not extend these measurements to the Embench suite.


We compare absolute execution times across runtimes for each benchmark snippet of the Embench suite as it represents real-world tasks that come up in embedded software. While Embench does specify a procedure for calculating a score, we decided not to use it as it directly depends on the execution time. We also used the Embench suite to measure the peak dynamic memory usage attained for each benchmark on the Pico2W. We measured it as the sum of the peak size of the heap, the \texttt{.data} section and the \texttt{.bss} section. We only measured it on the Pico2W for two reasons: firstly, we expect results to stay consistent across hardware architecture; and secondly, we wanted to evaluate every runtime on all benchmarks which is only possible on the Pico2W as some runtimes cannot complete some benchmarks on other hardware precisely because of Out-of-Memory errors. 

\subsection{Micro-benchmarks Results}

As a first comparison, Table \ref{tab:coremark-res} shows the CoreMark scores we measured for each runtime on the four boards that we used. We immediately observe that the performance of Wasm-interpreter and Wasefire are comparatively extremely low. Thus, in the next figures, we will leave out the measurements for those two runtimes as  they would otherwise significantly reduce readability. For example, on the Pico2W we measured Wasm-interpreter to be up to 15 times slower than the next slowest runtime and Wasefire to be between 3 and 4 times slower than Wasm-interpreter. The raw execution times can be found alongside the source code in this paper's contributions.

\begin{figure}[ht]
    \centering
    \includegraphics[width=0.99\linewidth]{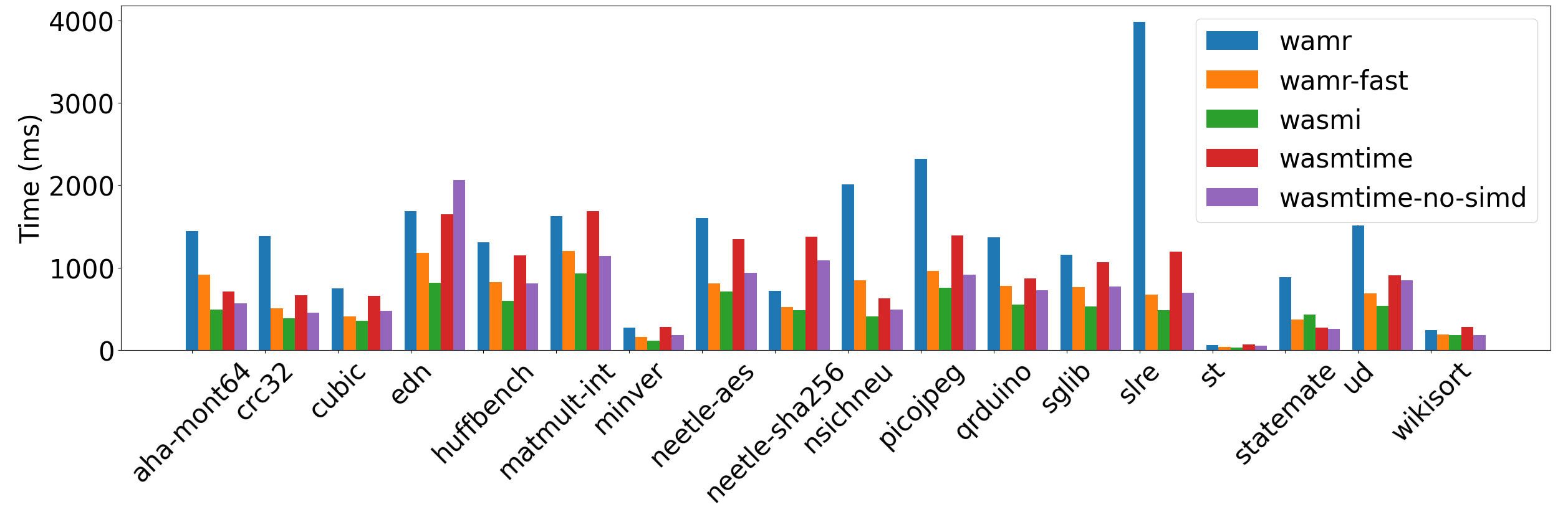}
    \caption{Absolute execution times of each benchmark in the embench suite on the Pico2W}
    \label{fig:timings-pico2}
\end{figure}

\begin{figure}[ht]
    \centering
    \includegraphics[width=0.99\linewidth]{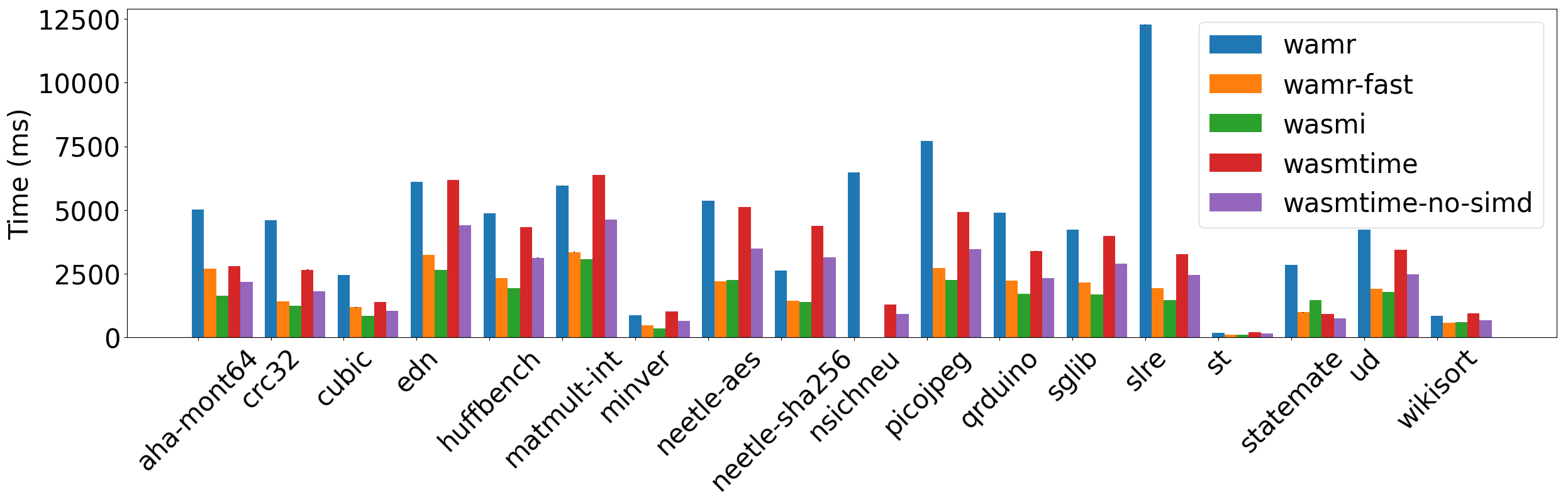}
    \caption{Absolute execution times of each benchmark in the embench suite on the nRF52840DK}
    \label{fig:timings-nrf52840}
\end{figure}

\begin{figure}[ht]
    \centering
    \includegraphics[width=0.99\linewidth]{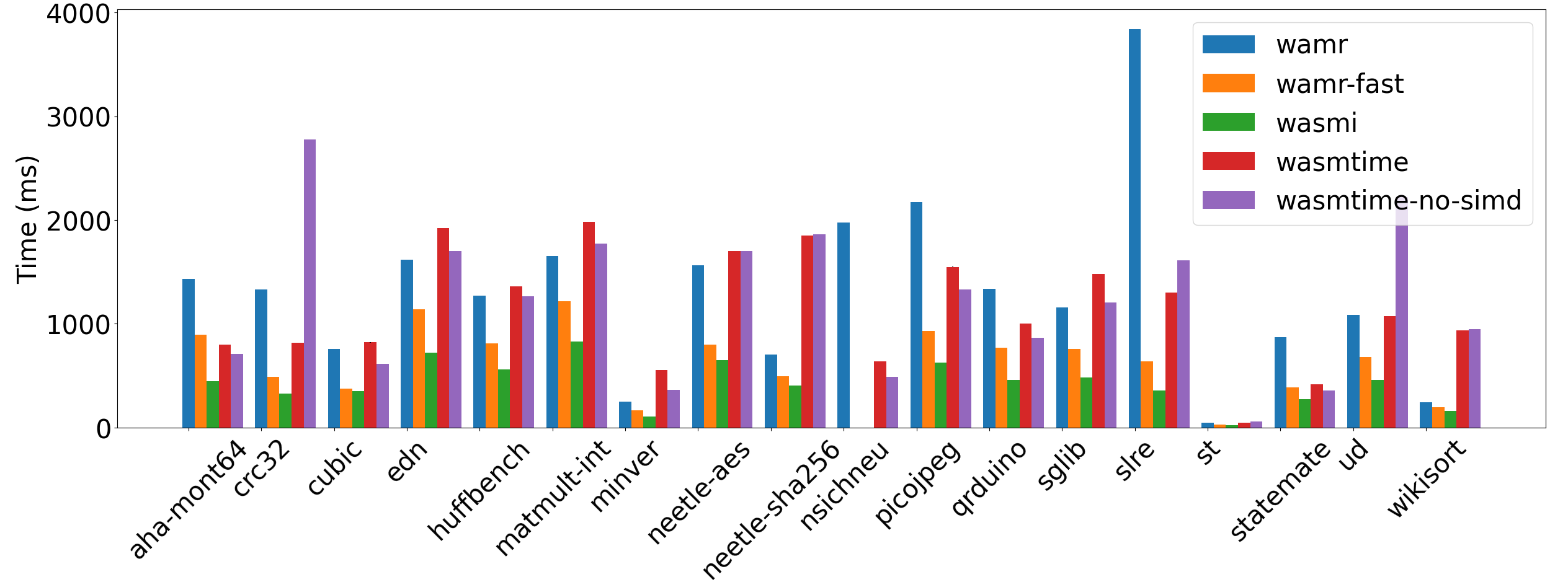}
    \caption{Absolute execution times of each benchmark in the embench suite on the ESP-WROOM-32}
    \label{fig:timings-esp32}
\end{figure}

\begin{figure}[ht]
    \centering
    \includegraphics[width=0.99\linewidth]{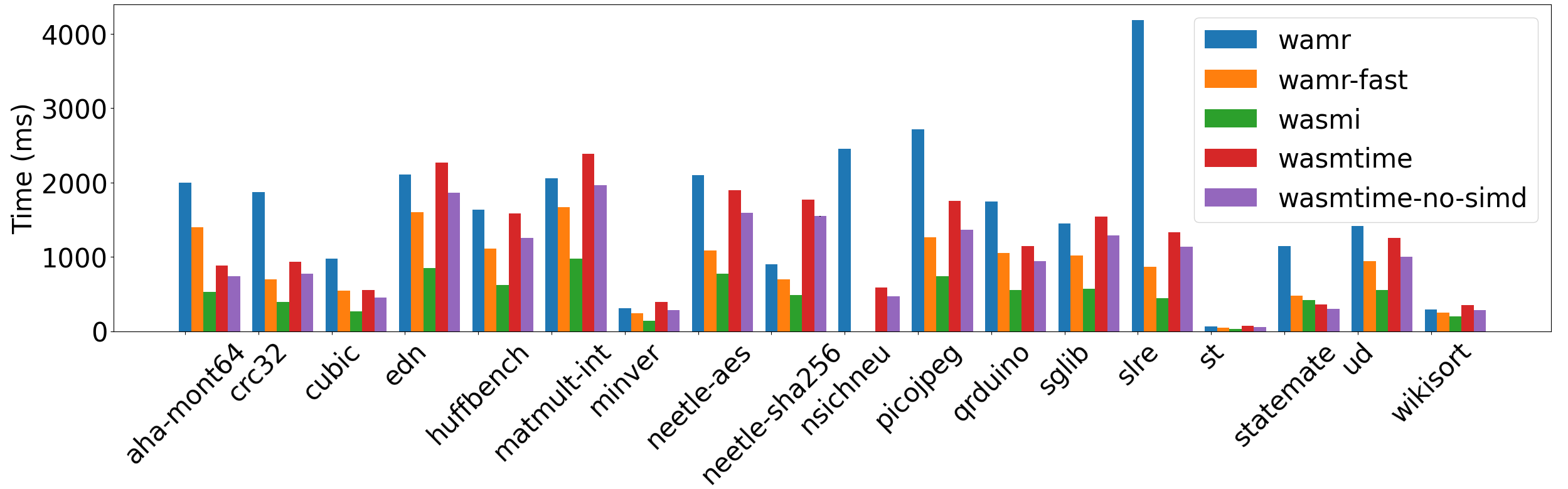}
    \caption{Absolute execution times of each benchmark in the embench suite on the ESP32-C6}
    \label{fig:timings-esp32c6}
\end{figure}

In nearly every time measurements, the geometric standard deviation was smaller than $0.1\%$. This shows that all studied WebAssembly runtimes have very consistent and reliable execution times for each benchmark snippet.
Figures \ref{fig:timings-pico2}, \ref{fig:timings-nrf52840}, \ref{fig:timings-esp32} and \ref{fig:timings-esp32c6} show the absolute execution times for the Pico2W, the nRF52840, the ESP-WROOM-32 and the ESP32-C6 respectively. 

\begin{figure}[ht]
    \centering
    \includegraphics[width=0.99\linewidth]{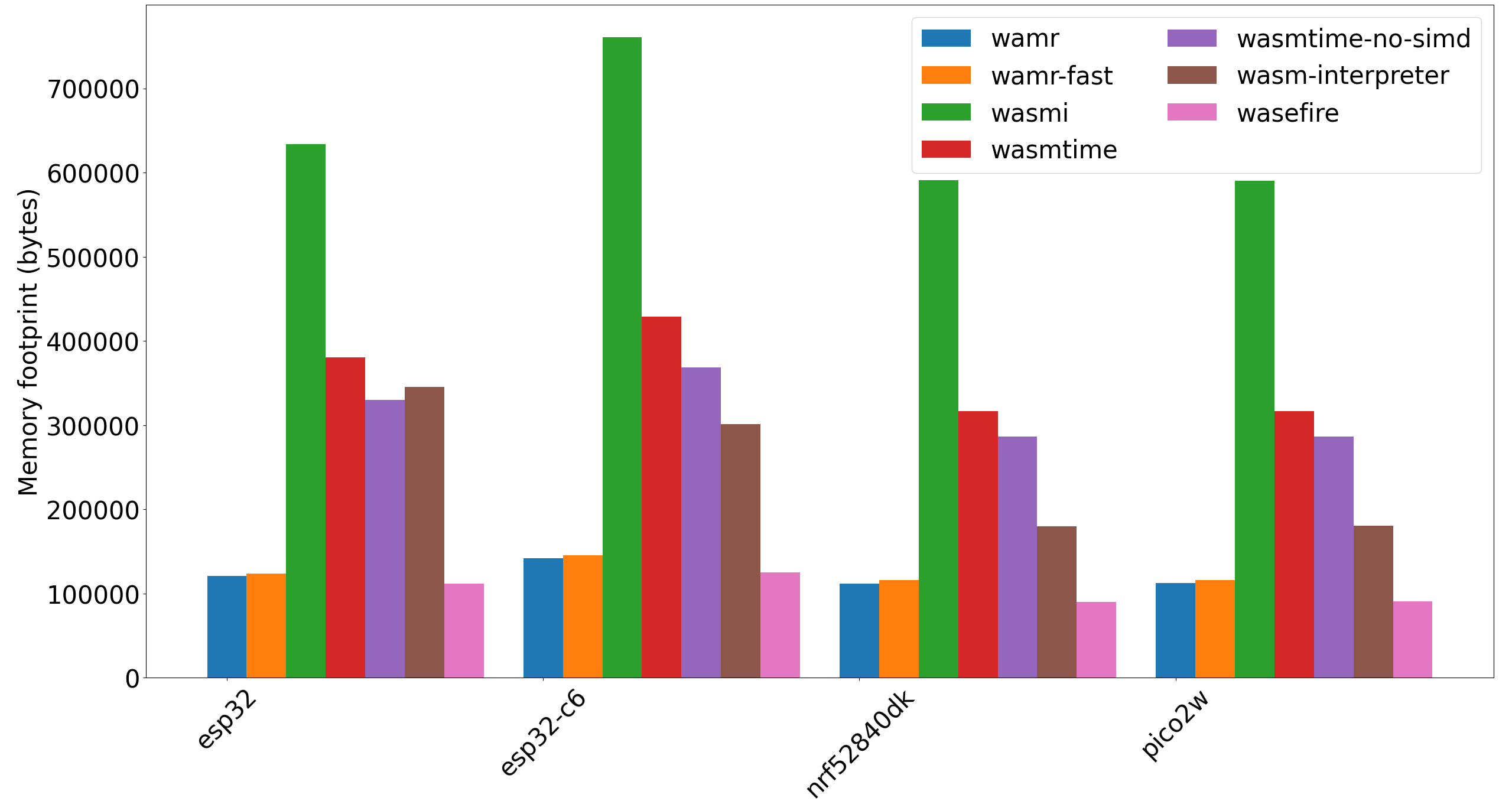}
    \caption{Flash memory size for CoreMark benchmarks on ESP32 (Xtensa), nRF52840, RaspberryPi Pico 2 (Cortex-M) and ESP32-C6 (RISC-V).}
    \label{fig:static-size}
\end{figure}

Static memory usage (Flash footprint) was measured for each runtime on the four boards on the CoreMark benchmark. This is shown in Figure \ref{fig:static-size}. 

\begin{figure}[ht]
    \centering
    \includegraphics[width=0.99\linewidth]{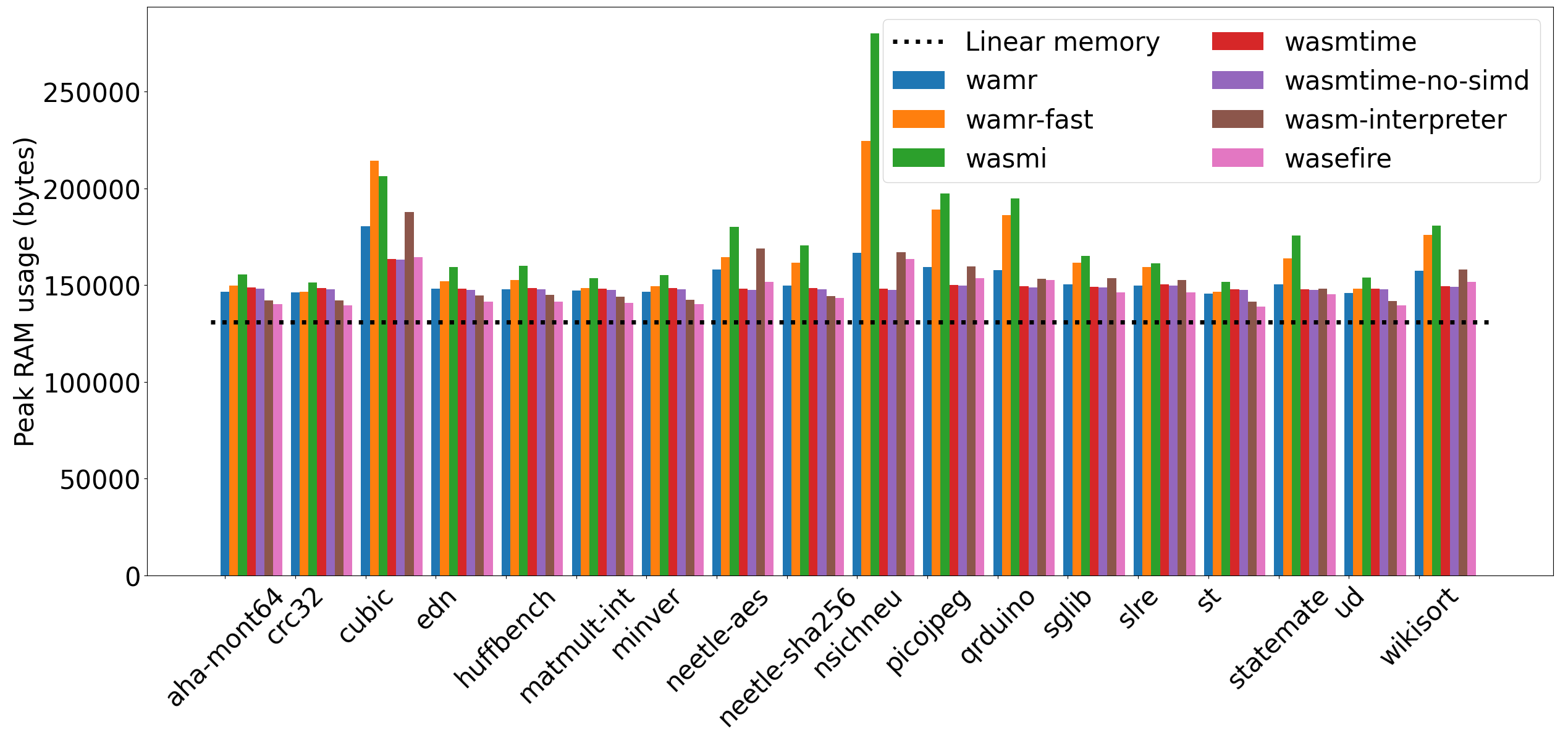}
    \caption{Peak RAM usage for the Embench suite on the Pico2W}
    \label{fig:peak-ram}
\end{figure}

Peak dynamic memory (Peak RAM footprint) usage is shown for the Embench suite on the Pico2W on Figure \ref{fig:peak-ram}. The dotted line corresponds to the two pages of Wasm linear memory that were used for all of the snippets. Anything above that line is runtime dependent.

\begin{table}[ht!]
    \caption{CoreMark scores on Cortex-M, RISC-V and Xtensa.}
    \label{tab:coremark-res}
    \centering
    \scriptsize
    \begin{tabular}{rcccc}
		\toprule
                          & RP2350 & ESP-WROOM-32 & nRF52840 & ESP32-C6 \\
		\midrule
         WAMR             & 6.6    & 6.7          & 1.8      & 5.3      \\
         WAMR Fast        & 10.8   & 11.0         & 3.8      & 8.0      \\
         Wasmi            & 15.5   & 17.3         & 4.8      & 14.5     \\
         Wasmtime         & 13.2   & 11.6         & 3.5      & 9.9      \\
         Wamstime No SIMD & 18.6   & 11.7         & 5.1      & 12.3     \\
         Wasm Interpreter & 0.6    & 0.5          & 0.2      & 0.6      \\
         Wasefire & 0.2 & 0.3           & 0.06     & 0.3      \\
		\bottomrule
         
    \end{tabular}
\end{table}

\subsection{Discussion}
Our first finding is that Wasefire and Wasm-interpreter are far too slow and offer limited compensations to justify using them in our prototype. These runtimes are quite new compared to the other candidates and have a compeling flash and RAM footprint so we will follow their development closely but we won't consider them further in the context of this work.

On the other side of the execution speed spectrum, our measurements demonstrate that Wasmi is overall the fastest interpreter on the various hardware we tested. However, Wasmi is also the runtime which leads to the largest code sizes, as showcased by our memory usage measurements. In terms of static memory footprint, WAMR's two runtime variants are significantly better than the rest of the competition. However, when looking at the dynamic memory consumption, we observe that, compared to the Wasmtime runtimes, WAMR peaks significantly higher on some benchmarks -- to the point where WAMR's fast interpreter uses too much memory on the \emph{nsichneu} benchmark on all hardware but the Pico 2.

\subsection{Wasm Runtime Choice Rationale}
Based on our measurements, there is no silver bullet. However, as our goal is to balance overall RAM footprint with execution speed and static (Flash) memory usage, we chose Wasmtime as a reasonable compromise. Indeed, our measurements indicate that Wasmtime is fast while not bloating the binary as much as Wasmi. Wasmtime also  consistently has one of the smallest RAM footprint. We also choose to use the Wasmtime variant that disables SIMD (Single Instruction, Multiple Data) capabilities as it proves to be faster in most cases while always leading to smaller binaries than its counterpart. It also marginally decreases the peak dynamic memory usage. This phenomenon is surprising and difficult to explain, but does match observations already made in previous work~\cite{no-simd-perf}.

Aside from the quantitative aspects we covered above, Wasmtime also provides several qualitative advantages. First, Wasmtime is implemented in Rust, which works well with our targeted embedded Rust environment (i.e. the embedded platform provided by Ariel OS). This characteristic allows to reduce the number of different toolchains that one has to combine in the end. Second, Wasmtime also follows more closely the latest developments in WebAssembly standardization than any other runtime. For example, Wasmtime is one of the few, if not the only, runtime fit for embedded systems supporting the WebAssembly Component Model \cite{componentmodel} which offers a standard way to communicate high-level logic between Wasm capsules and/or with their host OS. Wasmtime also supports the WebAssembly Custom Page Size proposal \cite{custompagesize} which allows capsules to request linear memory under the default of 64KiB which is often much more than needed for embedded applications and can be a pain point when dealing with several independent capsules that need to have their own linear memory. Last but not least, Wasmtime is also, to our knowledge, one of the few runtimes that allows for some asynchronous execution of capsules using the Rust async/await paradigm.

\section{Design of the \emph{treVM} Prototype}

In this section, we give a high-level overview of our tiny rust embedded virtual machine (\emph{treVM}) platform. What we mean here by the term generic is that \emph{treVM} is designed so that it can deploy and run Wasm capsules on top of a large variety of heterogeneous MCUs. Also, not only can \emph{treVM} capsules host highly customizable, arbitrary code, but also capsules can be securely updated over the network, on demand, on devices already deployed in the field.

\subsubsection*{Embedded Software Stack Basis}

\ \\A typical software stack is assumed to be embedded on networked microcontroller-based hardware such as we target (e.g. wireless sensors, actuators etc.). Usually, this software stack includes an RTOS which (i) provides some level of hardware abstraction, (ii) includes basic functionalities including sensor/actuator drivers, as well as (iii) bundles a default network stack to communicate securely over the network. 

In our design, we use Ariel OS~\cite{frank2025ariel} as basis RTOS, and the 6LoWPAN network stack (IPv6, UDP, CoAP), secured with protocols such as OSCORE and EDHOC (see \cite{tschofenig2019cyberphysical}) for the default network subsystem (which can work on various link layers including IEEE 802.15.4, BLE, serial-USB, Ethernet, WiFi...). Conveniently, Ariel OS bundles Rust implementations for all of the necessary subsystems, on top of which we can then build \emph{treVM}.
\ \\

\subsubsection*{Capsules \& System Calls Bindings}
\ \\ As motivated in the previous sections, we use  Wasmtime as Wasm runtime for \emph{treVM} capsules. 

By default, capsules cannot access any part of the host OS. In some cases however, it is desirable to allow the capsules to access a predefined subset of the functionality provided by the host OS, for example sensor readings, random number generation or logging interfaces. To allow for this, we leverage the WebAssembly Component Model \cite{componentmodel} which allows us to declare interfaces between the host and the capsules. The interfaces are declared using the WIT (WebAssembly Interface Type) language and designed starting from the OS. We tried to mimic the OS's behavior as much as possible to make capsule development easier for users of the OS while keeping an easy-to-understand structure. Listings \ref{listing:sensor} shows how a capsule can interact with the sensor API provided by Ariel OS.


\begin{listing}[b]
\includegraphics[width=0.99\linewidth]{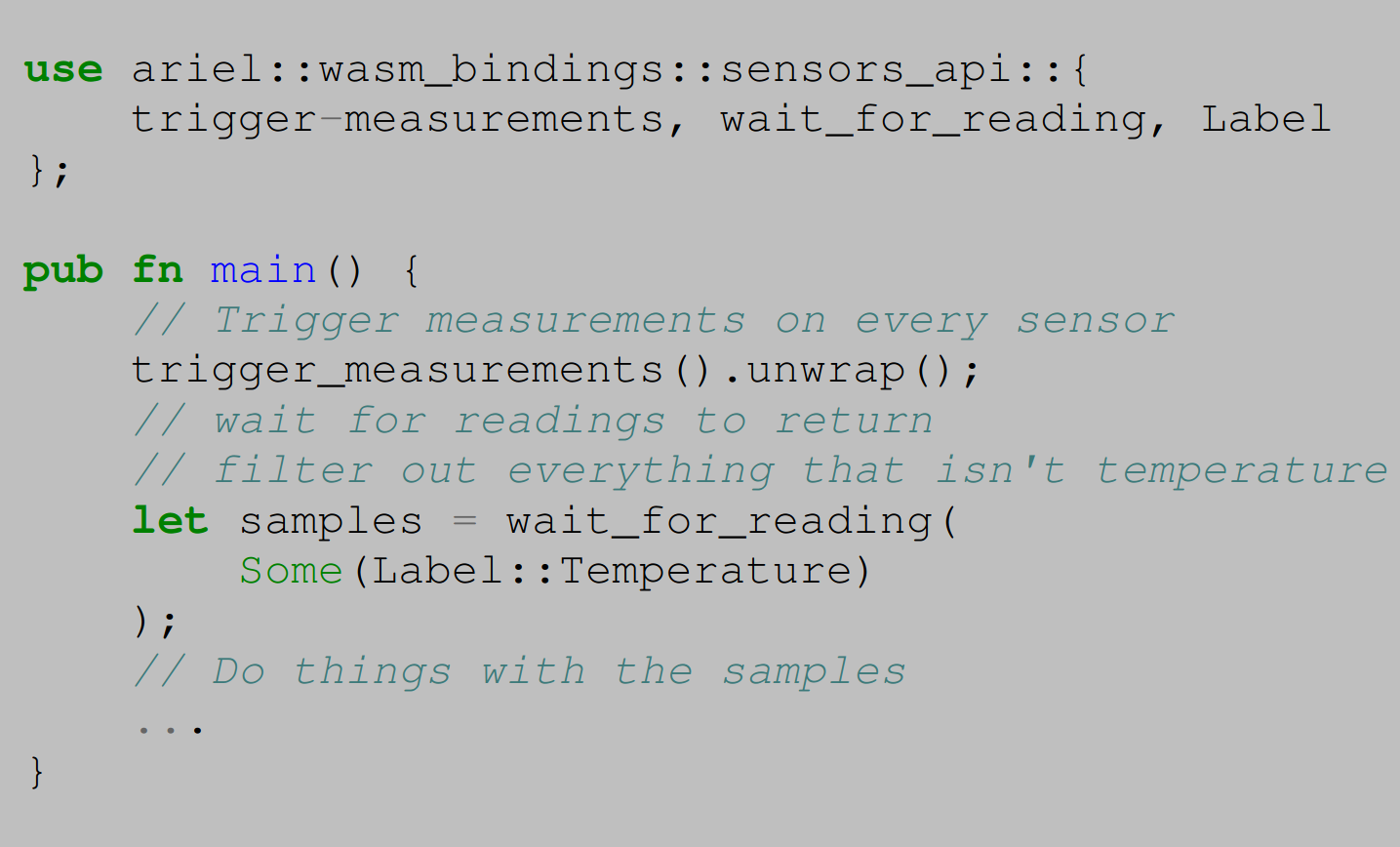}
\caption{Usage of the Sensors API in the capsule}
\label{listing:sensor}
\end{listing}

\begin{figure}[t]
    \centering
    \includegraphics[width=0.99\linewidth]{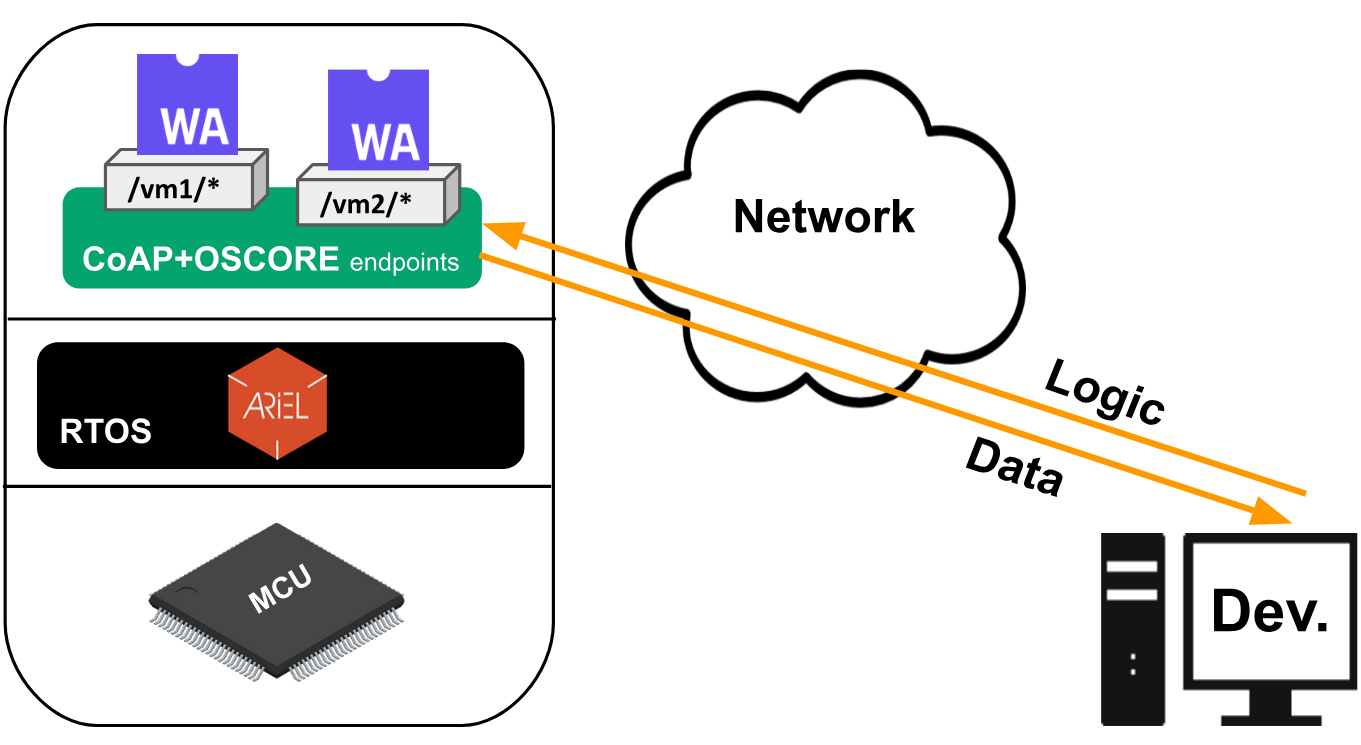}
    \caption{CoAP endpoints with \textit{treVM} in Ariel OS.}
    \label{fig:trevm-coap}
\end{figure}

\begin{figure}[t]
    \centering
    \includegraphics[width=0.58\linewidth]{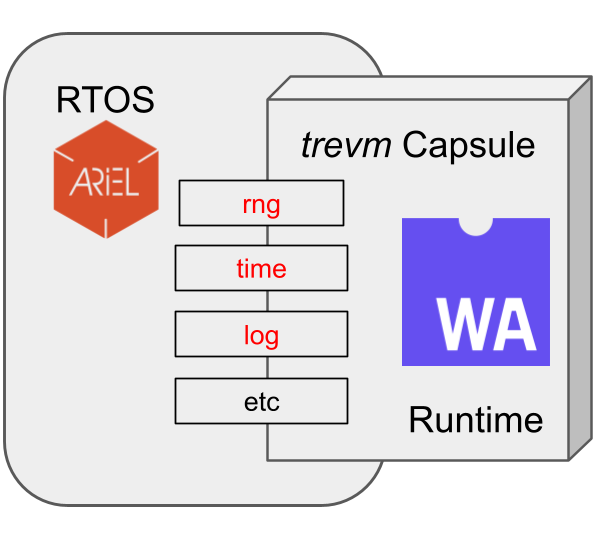}
    \caption{System calls for \textit{treVM} in Ariel OS.}
    \label{fig:trevm-syscalls}
\end{figure}

Interfaces are grouped by functionality, leading to the architecture shown in figure \ref{fig:trevm-syscalls}. 
The system calls we implemented in our prototype so far (rng, timer, log, and high-level sensor APIs) serve as blueprint for future, additional system calls, to be implemented along the way as required by more diverse use cases.

\subsubsection*{Interacting with Capsules over the Network}

\ \\ An important design aspect of \emph{treVM} is the way it integrates capsules within a CoAP server, secured with OSCORE \cite{rfc8613}. As a reminder, such a network stack~\cite{tschofenig2019cyberphysical} is typically provided as basis for secure communication on various RTOS, including on Ariel OS, the basis we use in our design.

Once initially deployed, interaction with the capsules can happen through the provided CoAP server. We distinguish two levels of interaction for the capsules (i) ephemeral capsules, and (ii) persistent capsules.

{\bf Ephemeral capsules} are sent over the network via CoAP, launched, run to completion and then may send data back to the sender, typically as part of the same CoAP exchange, assuming they are short-running  (otherwise data can be sent back through other means). 

\begin{figure}[t]
    \centering
    \includegraphics[width=0.99\linewidth]{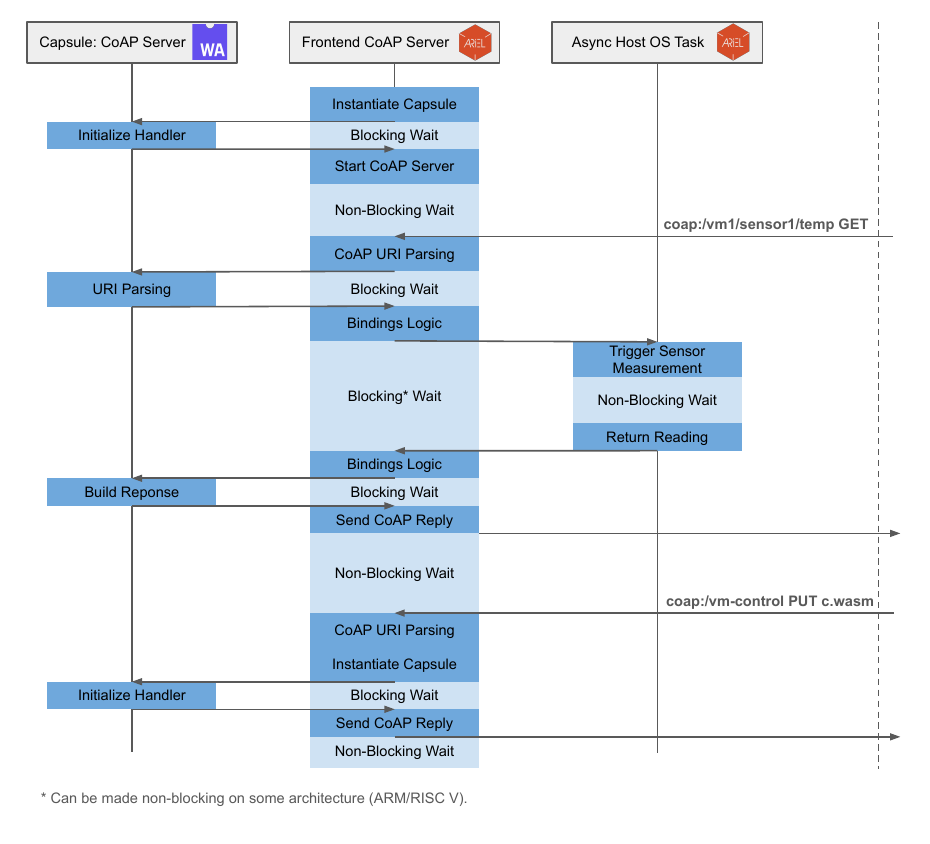}
    \caption{Cinematic of the usage and update of persistent capsules in \emph{treVM}.}
    \label{fig:trevm-cinematic}
\end{figure}

{\bf Persistent capsules}, on the other hand, are not only deployed with CoAP but also given a dedicated top-level URI path (for instance \emph{/vm1/*}) and are thereafter reachable with CoAP requests which match this path.  Such capsules can be queried multiple times, until they are updated or otherwise terminated by the CoAP server. Furthermore, requests to any resource further down their dedicated URI path (for instance \emph{/vm1/sensor1/temp}) are handled \emph{within} the capsule. Hence, this specific part of the logic remains entirely flexible/updatable, independently of the rest of the embedded software stack. However, this comes at the price of slightly bigger capsules in that case (as part of the CoAP handler logic must be implemented in the capsule). Figure \ref{fig:trevm-coap} shows the case where two \emph{treVM} capsules are deployed on the same MCU, and reachable over the network using URIs \emph{/vm1/} and \emph{/vm2/}. Figure \ref{fig:trevm-cinematic} showcases the cinematic of an example scenario with a capsule that gives access to sensor readings. At startup, the OS level CoAP server instantiates and initializes the capsule. After that a client asks for a reading by sending a GET request for the \texttt{vm1/sensor1/temp} resource. Following the URI parsing logic, this request reaches the capsule which uses its bindings with the host OS to gather sensor readings. The readings are then used to build the response to the GET request inside of the capsule which is then handed up to the OS level server which deals with the return path to the client. Later the capsule is updated via a PUT request at \texttt{/vm-control}. The OS level CoAP server uses the payload to instantiate a new capsule and return the appropriate status code.
\ \\
\subsubsection*{Security \& Fault-Tolerance Considerations}
\ \\ In \emph{treVM} we distinguish between two origins for crashes or other faulty behaviour regarding capsule execution. 
\ \\ \indent If the capsule contains valid Pulley Bytecode, since the WebAssembly specification \cite{WebAssemblyCoreSpecification1} and the security policy of our runtime of choice Wasmtime \cite{wasmtimesecurity} offer strong guarantees against sandbox escapes, only malfunctioning or malicious business logic like crashes or infinite loops can cause faulty behaviours. Wasmtime has guard-rails against both options, as do most other runtimes, in the forms of a fuel mechanism to cap the number of instructions that a capsule can run and the ability to catch capsule-side crashes using the trap infrastructure built into WebAssembly.

This leaves the need to prevent capsules containing invalid bytecode from reaching our micro-controllers. To achieve this, we rely OSCORE to ensure request authentication and integrity and on the fact that turning arbitrary input into Pulley bytecode is a safe operation that only outputs valid bytecode~\cite{precompile-safety}. This means that we can ensure bytecode validity by wiring update requests through a Wasm to Pulley translator module which would be the only entity authorized to update the capsules on the micro-controller.
\section{Evaluation of \textit{treVM}}
In this section, we evaluate our prototype \emph{treVM} implementation using the following metrics:
\begin{itemize}
    \item The total flash size of the firmware before any updates
    \item The peak RAM usage after doing at least one update
    \item The size of the capsule
\end{itemize}
We conducted this evaluation on a commonly available board: the nRF52840DK, which is based on an Arm Cortex-M4 micro-controller. We used the software toolchain specified in Section~\ref{sec-setup}. For the network stack, we used Ariel OS integration of smolTCP including CoAP and OSCORE, running on top of Ethernet-over-USB as the link layer (note, nevertheless, that other link link layers are also supported, including Bluetooth LE or WiFi for instance). Using the previous sections' terminology, we measured the performance of treVM for different the following use cases:
\begin{itemize}
    \item (i)  Ephemeral capsule without bindings
    \item (ii) Ephemeral capsules with bindings
    \item (iii) Persistent capsules without bindings
    \item (iv) Persistent capsules with bindings
\end{itemize}
Capsules without bindings implement simple logic with linear complexity (the Fibonacci suite using a 'for' loop). Capsules with bindings interact with a sensor by triggering the measurement, waiting for the reading, adding random noise to the reading and logging the result. All Wasm capsules are provisionned with a linear memory of 32768 bytes lowered from the standard minimum of 65536 bytes (the latter is the usual WebAssembly custom page size as specified in \cite{custompagesize}). Table \ref{tab:measure-trevm} shows our measurements. Note that at the bottom of this table, for reference, we also added measurements for a vanilla example that implements a minimal CoAP server over Ariel OS. The treVM Flash and RAM footprints in columns 3-4 are estimates. The RAM footprint is calculated as roughly the sum of the linear memory of the capsule and the binary size of the capsule. The latter needs to be considered because, when updating, the entire capsule needs to be put in RAM. The flash footprint is based on numbers we derived from \emph{cargo-bloat}\cite{cargobloat}. This method only uses the \texttt{.text} sections of the binary and thus does not measure precisely how much of the total \texttt{.data}/\texttt{.rodata} can be attributed to \emph{treVM}. Nevertheless, these estimations give a good ballpark already, and numbers in columns 2, 5 and 6 are actual measurements.

\begin{table}[ht]
    \caption{Resource footprint per \emph{treVM} capsule type (in kiloBytes).}
    \label{tab:measure-trevm}
    \scriptsize
    \centering
    \begin{tabular}{|r|c|c|c|c|c|}
        \hline
        Capsule              & Capsule  & Flash           &  RAM           & Flash  & RAM    \\
        Type                 & Size  & \emph{treVM}    & \emph{treVM}   & total  & total  \\
        \hline
        Ephemeral w/o bind.  & 9.6          & $\approx$208  & $\approx$40  & 560  & 60   \\ 
        \hline 
        Ephemeral            & 28.8         & $\approx$228  & $\approx$60  & 608  & 94   \\      
        \hline 
        Persistent w/o bind. & 34.5         & $\approx$212  & $\approx$70  & 600  & 117  \\
        \hline 
        Persistent           & 43.2         & $\approx$230  & $\approx$80  & 626  & 130  \\ 
        \hline
        \hline
        Minimal CoAP Server  & N/A            & N/A             & N/A            & 208  & 29    \\
        \hline
    \end{tabular}
\end{table}

Overall, we find that adding \emph{treVM} on top of an embedded Rust OS with a secure CoAP network stack requires more than doubling the Flash and RAM footprint for a simple capsule. This kind of ratio is not surprising. For instance, authors of prior work (see \cite{zandberg2022femto}  Table 1) reported that using a Wasm3 runtime for an equivalent ephemeral capsule without bindings on top of a common RTOS written in C (namely RIOT) led to +120\% Flash size. In \cite{zandberg2022femto}, authors also find that their capsule requires slightly more additional RAM (53KiB) compared to the 40KiB we observe for \emph{treVM} (note that we discount 32KiB from their 85KiB measurement, as they have used 64KiB pages, contrary to our 32KiB pages).
\section{Discussion}
In this paper we have demonstrated the feasibility of a Rust-based embedded software hosting Wasm software capsules, sandboxed and securely updatable over the network, applicable on a wide variety of commonly-available IoT devices with RAM and Flash memory budgets in the order of hundreds of kilobytes 
(such as RaspberyPi Pico, nRF52840DK or various ESP32 boards). This  indicates that \emph{treVM} is suitable for versatile use cases and, in particular, that \emph{treVM} does not require specialized hardware, and thus minimizes upfront costs. 

{\bf Regarding Flash footprint}, our measurements indicate that Flash size is bigger than a rough C equivalent found in prior work such as in~\cite{zandberg2022femto}. This phenomenon is well known from the literature~\cite{ayers2022tighten-rust}. However, on the one hand, the sizes still fit on most of the newest microcontroller hardware. On the other hand, while not explored in our paper, we expect Flash binary size of Rust firmwares will gradually be optimized over time, as C binaries have themselves been optimized over the last decades.

{\bf Regarding reducing RAM footprint}, we mentioned in our \emph{treVM} evaluation that our capsules have a linear memory of 32768 bytes. This number was chosen to highlight our use and compliance with the WebAssembly custom page size proposal, but does not reflect the actual need of the capsules. Considering the relative simplicity of the evaluated capsules, it is not unreasonable to expect 16 KiB, or maybe even less, to suffice. Implementing this optimization would directly reduce the overall RAM usage. Optimizing the stack size requirements of the capsule can also be attempted in conjunction, which would unlock smaller linear memory requirement. As both the stack size and linear memory size are directly controlled through compile-time arguments, pin-pointing acceptable sizes is a relatively low-hanging fruit. Hence, the numbers we report are higher bounds, rather than lower bounds.

{\bf Regarding reducing capsule size}, this task is more challenging. First of all, it is important to recall that the capsules we are talking about are not pure WebAssembly bytecode but Pulley Bytecode compiled for Wasmtime's Pulley interpreter. This bytecode is more efficient to interpret than Wasm bytecode but it takes more space and is packaged in an ELF file that can itself be optimized. Some sections like \texttt{.strtab} and \texttt{.symtab} contains data that is not used in any way by Wasmtime and can be removed without consequences. Other like \texttt{.wasmtime.engine} and  \texttt{.wasmtime.bti} contain data regarding the expected run-time configuration which is double-checked during capsule initialization. Removing these last section is probably acceptable in our context, where communications are secured and only authenticated users can upload and update capsules.   

\section{Future Work}
Currently, our implementation relies heavily on the WebAssembly Component Model~\cite{componentmodel}. We expect this model to be supported by multiple different runtimes in the future. However, in the meantime, we plan to develop a variant of our process that is adapted to plain WebAssembly. This would facilitate switching between different runtimes in the 'backend', to explore a larger variety of trade-offs. 

We plan to keep up closely with state-of-art embedded Rust, which is a fast-moving target.  This is vital in view of exploiting more optimization opportunities to downsize our firmware and the capsules. We expect this to happen naturally over time as Rust matures. 

We note that the landscape of Wasm runtimes is also a fast-moving target. New runtimes are published frequently, and existing runtimes are also being updated quite often. Our micro-benchmarks will need to be updated regularly to reflect this evolution. By publishing the source code of both \emph{treVM} and the benchmarks, we expect that this task will be facilitated and spread throughout the  community of embedded software researchers. Wasefire, Wasm-interpreter and Myrmic are all interesting targets in that regard, as they are either soon to be open-sourced or actively developed and subject to important changes fairly often. 

Finally, adding more interfaces between the capsules and the OS is a priority to facilitate the use of \emph{treVM} for more applications. Interfaces to common network stacks (such as Bluetooth-Low-Energy or LTE-M) must be expanded and fully fleshed out. In order to unlock other use cases for \emph{treVM}, yet other interfaces could allow capsules to monitor additional system information, such as power consumption.
\section{Conclusion}

In this paper, we address the joint rise of two trends that impact software on microcontroller-based embedded systems. On the one hand, Rust is becoming the language of choice for programming such devices. On the other hand, small WebAssembly runtimes emerge as applicable components to sandbox parts of the logic running on embedded devices. We thus performed a reality check combining these approaches. We first provided a comprehensive comparative performance evaluation of the main WebAssembly runtime alternatives targeting microcontrollers. Then, based on these results, we designed \emph{treVM}, a generic Rust embedded software platform that enables both the execution of arbitrary WebAssembly code capsules on heterogeneous microcontrollers (Cortex-M, RISC-V, ESP32) and their secure deployment over the network to reprogram devices up and running in the field. The applicability of \emph{treVM} is augmented by harnessing the versatile hardware abstraction provided by a general-purpose Rust RTOS (Ariel OS). We demonstrate the feasibility of \emph{treVM}, and measure its performance on commercially available microcontroller hardware. We also published the source code of its implementation. As such, \emph{treVM} provides a powerful continuous deployment (CI/CD) tool for next-generation Rust embedded software platforms.

\section*{Source Code Availability}
The source code for \emph{treVM} is  maintained at \href{https://github.com/ariel-os/trevm}{\textbf{https://github.com/ariel-os/trevm}}. The code used for benchmarks is available at \href{https://github.com/anlavandier/trevm-dcoss2026}{\textbf{https://github.com/anlavandier/trevm-dcoss2026}} and at \href{https://github.com/anlavandier/ariel-runtime-size-comparisons}{\textbf{https://github.com/anlavandier/ariel-runtime-size-comparisons}}.

\section*{Acknowledgements}
The work described in this paper was in parts financed by France 2030 via the PEPR Future Networks project FITNESS, and the PTCC project PQ-OTA.

\bibliographystyle{IEEEtran}
\bibliography{bibliography,bibliographyBB}

\end{document}